\documentclass[aps,prb,twocolumn,groupedaddress]{revtex4}

\usepackage{graphicx, color}
\usepackage{dcolumn}
\usepackage{bm}
\usepackage{amsmath}
\usepackage{amssymb}
\usepackage{verbatim}
\usepackage{appendix}
\usepackage{soul}
\usepackage{gensymb}
\usepackage{siunitx}
\usepackage{subcaption}
\newcommand{\nv}{NV$^-$ }
\newcommand{\cthirt}{$^{13}$C }
\newcommand{\trec}{$t_{rec}$ }
\newcommand{\fsat}{$f_{MW}$ }
\newcommand{\nvn}{NV$^0$ }
\newcommand{\mspm}{m$_s=\pm1$ }
\newcommand{\msz}{m$_s=0$ }
\newcommand{\msp}{m$_s=+1$ }

\newcommand{\tdnp}{T$_{DNP}$ }
\newcommand{\tone}{T$_1$ }
\newcommand{\tonecth}{T$_{1,^{13}\mathrm{C}}$ }

\newcommand{\pone}{N$_S$ }
\newcommand{\ttwo}{T$_2$ }

\begin{document}



\title{Optically-pumped dynamic nuclear hyperpolarization in $^{13}$C enriched diamond}



\author{Anna J. Parker}
\affiliation{Department of Chemistry, University of California, Berkeley, California 94720, United States}
\affiliation{Materials Sciences Division, Lawrence Berkeley National Laboratory, Berkeley, California 94720, United States}
\author{Keunhong Jeong}
\affiliation{Department of Chemistry, University of California, Berkeley, California 94720, United States}
\affiliation{Materials Sciences Division, Lawrence Berkeley National Laboratory, Berkeley, California 94720, United States}
\author{Claudia E. Avalos}
\affiliation{Department of Chemistry, University of California, Berkeley, California 94720, United States}
\affiliation{Materials Sciences Division, Lawrence Berkeley National Laboratory, Berkeley, California 94720, United States}
\author{Birgit J. M. Hausmann}
\affiliation{Department of Chemistry, University of California, Berkeley, California 94720, United States}
\affiliation{Materials Sciences Division, Lawrence Berkeley National Laboratory, Berkeley, California 94720, United States}
\author{Christophoros C. Vassiliou}
\affiliation{Department of Chemistry, University of California, Berkeley, California 94720, United States}
\affiliation{Materials Sciences Division, Lawrence Berkeley National Laboratory, Berkeley, California 94720, United States}
\author{Alexander Pines}
\affiliation{Department of Chemistry, University of California, Berkeley, California 94720, United States}
\affiliation{Materials Sciences Division, Lawrence Berkeley National Laboratory, Berkeley, California 94720, United States}
\author{Jonathan P. King}
\affiliation{Department of Chemistry, University of California, Berkeley, California 94720, United States}
\affiliation{Materials Sciences Division, Lawrence Berkeley National Laboratory, Berkeley, California 94720, United States}
\email{jpking@berkeley.edu}

\date{\today}

\begin{abstract}
We investigate nuclear spin hyperpolarization from nitrogen vacancy centers in isotopically enriched diamonds with \cthirt concentrations up to 100\%.  $^{13}$C enrichment leads to hyperfine structure of the nitrogen vacancy electron spin resonance spectrum and as a result the spectrum of dynamic nuclear polarization. We show that strongly-coupled \cthirt spins in the first shell surrounding a nitrogen vacancy center generate resolved hyperfine splittings, but do not act as an intermediary in the transfer of hyperpolarization of bulk nuclear spins. High levels of \cthirt enrichment are desirable to increase the efficiency of hyperpolarizaiton for magnetic resonance signal enhancement, imaging contrast agents, and as a platform for quantum sensing and many-body physics.


\end{abstract}


\maketitle


The last decade has witnessed rapid strides in the development of quantum technologies based on atom-like defect centers in solids\cite{koehl11,widmann15}, such as the negatively-charged nitrogen vacancy (\nv) center \cite{Doherty2013}. The \nv is a system of six localized electrons in diamond with a total spin of 1, whose properties have drawn attention from various scientific fields. For example, optical initialization of the \nv spin state \cite{Manson07}, long electron spin coherence times exceeding 1 ms \cite{Balasubramanian09,christle15}, and optical spin state readout \cite{Doherty11} have made the defect a model platform for quantum information processing \cite{Yao12,Hensen15,klimov15}, simulation \cite{Cai13,Ajoy13b} and metrology \cite{Taylor08,Maze08,Balasubramanian08}.

Nearly all of these technologies require knowledge of and capitalization upon the interactions of the \nv center with nearby nuclear spins both in and external to the diamond. Tuning the \cthirt concentration by isotopic growth techniques \cite{Bar-Gill12} enables a variety of quantum technology schemes. For instance, in the limit of low nuclear concentrations ($\leq$ 1\%), \nv-\cthirt pairs can form quantum registers \cite{Dutt07,Neumann10,Reiserer16} with increased sensing resolution \cite{Laraoui13,Rosskopf16} and sensitivity \cite{Ajoy2016}. At slightly higher concentrations ($\sim$ 10\%), a single \nv center can be coupled to a number of \cthirt nuclei to form the node of a quantum information processor, allowing the indirect fast actuation and universal quantum control on the nuclear spins via the electronic qubit \cite{Khaneja07,Borneman12}. At high concentrations beyond 50\% and approaching 100\% where internuclear couplings become significant, hybridized nuclear spin states enable decoherence protected subspaces \cite{kalb17} where classical information can stored. This also forms a versatile system to study various condensed matter phenomena in the strongly dipolar coupled quantum networks \cite{Weimer13}, including notions of quantum transport \cite{Ajoy13}, localization and criticiality \cite{Yao14,Nandkishore15}, and Floquet many-body phases \cite{Khemani16,Else16,Choi17}. 

Additionally, integrated \nv-nuclear spin systems provide an exciting opportunity for long-standing Nuclear Magnetic Resonance (NMR) and Magnetic Resonance Imaging (MRI) technologies. While NMR and MRI are indispensable tools to the fields of chemistry, biology, engineering and medicine, their sensitivity relies on nuclear spin initialization (i.e. polarization), which at best reaches 10$^{-4}$ at room temperature. In stark contrast to the weak magnetization of nuclear spins, optical pumping hyperpolarizes the \nv spin state beyond thermal equilibrium at arbitrary temperature and over a wide range of magnetic fields \cite{Scott16}. As a result, a number of schemes for creating nuclear spin hyperpolarization with \nv centers based on traditional dynamic nuclear polarization (DNP) methods \cite{DNPbook} have been reported in the recent literature\cite{Fischer2013,Alvarez2015,King2010,Scheuer2016,Waddington2016,Rej2016,Chen2015,Chen2016,King2015}. These schemes propose the use of hyperpolarized $^{13}$C nuclei in diamond for use as MRI contrast agents\cite{Rej2015} as well as a platform for polarization transfer to external nuclei\cite{King2015} for enhanced magnetic resonance signal from arbitrary samples. In these cases, the low natural abundance (1.1\%) of $^{13}$C nuclear spins limits the efficiency of hyperpolarization and it is desirable to work with \cthirt enriched materials. 

Here, we report NV$^-$ DNP hyperpolarization of $^{13}$C enriched diamonds. Our methods result in significant \cthirt bulk polarizations approaching 0.1\% at approximately 0.5 T in a variety of samples, an enhancement of three orders of magnitude over thermal polarization. We show how isotopic enrichment imparts a complex structure to the electron spin resonance spectrum and corresponding DNP spectrum. The DNP spectra lend insight to the mechanism of polarization transfer, illustrating that \nv centers and the first shell of $^{13}$C spins behave as a strongly coupled system that transfers polarization directly to weakly coupled nuclear spins. These findings open the path to their use as efficient external polarizing agents, and for applications in quantum technologies.

 \begin{figure}[h!]
 \includegraphics[width=0.5\textwidth]{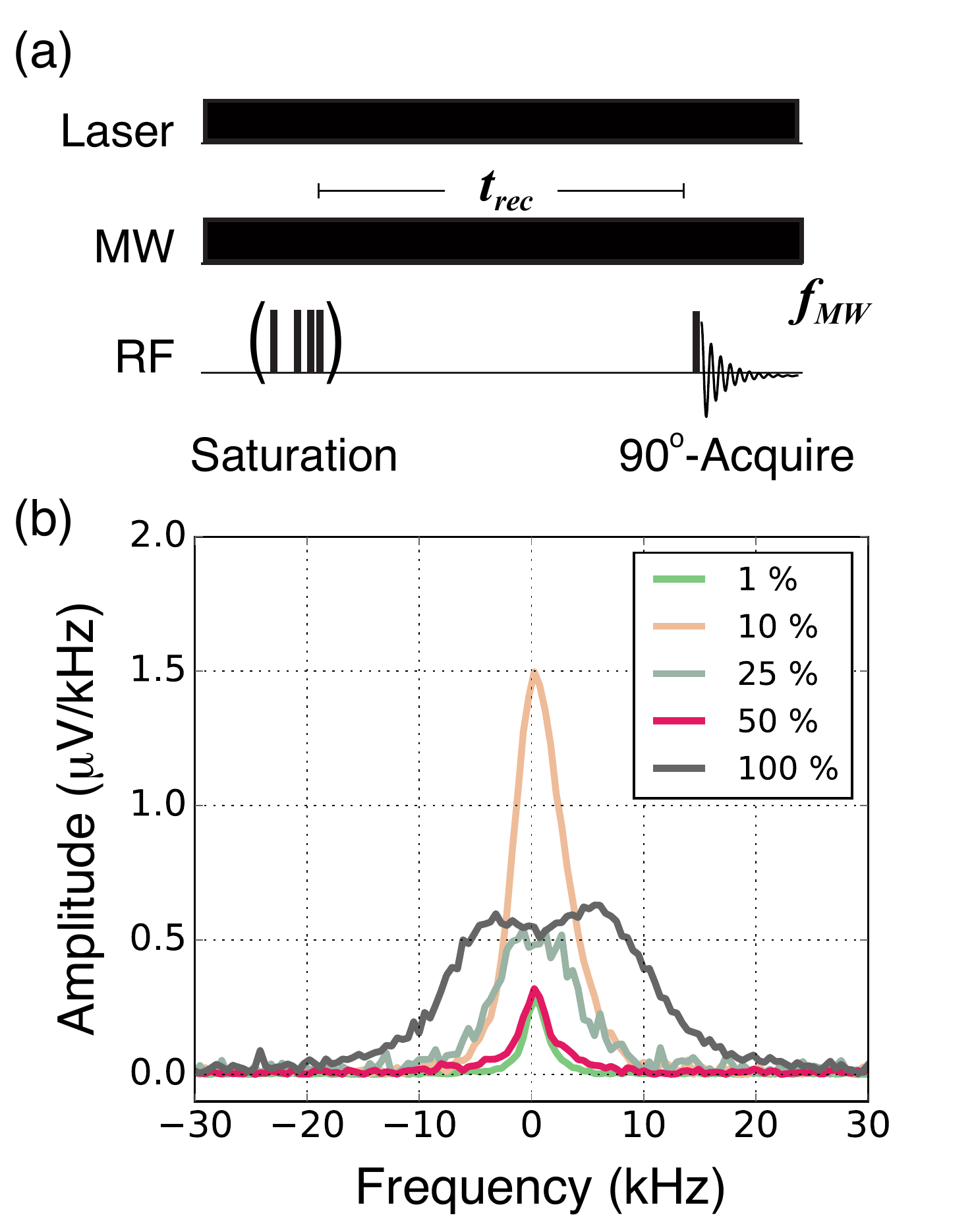}%
 \caption{\label{fig:dnp_exp} CW-DNP with \nv centers. A schematic for the CW-DNP experiment is shown in (a). The experiment begins with a series of 90\degree pulses to destroy \cthirt polarization, after which the \cthirt polarization builds under continuous optical and microwave excitation. The microwave frequency \fsat is swept to find the optimum frequency. After the \cthirt nuclear polarization builds during a recovery time \trec, a single 90\degree pulse is applied and the NMR signal is acquired. The spectra of the hyperpolarized \cthirt NMR resulting from optimizing \fsat and the polarization buildup time are shown in (b) for samples with a natural abundance (1.1\%) \cthirt and 10, 25, 50, and 100\% \cthirt enrichment. }
 \end{figure}

In this work we employ continuous-wave (CW) DNP to hyperpolarize \cthirt nuclei in a set of single crystal diamonds at a magnetic field of approximately 472 mT with the \nv crystal axis aligned along the magnetic field. A schematic of the DNP pulse sequence is given in Figure \ref{fig:dnp_exp}a. Each experiment begins with a set of 90\degree pulses to destroy any thermal \cthirt polarization before a recovery time $t_{rec}$ where \cthirt polarization builds under continuous optical and microwave irradiation. The laser optically pumps the \nv center to continually initialize its spin state, while microwave irradiation has the effect of transferring spin polarization between \nv and \cthirt spins, thus hyperpolarizing the \cthirt spins and producing enhanced NMR signals. The hyperpolarized \cthirt NMR spectra of each sample are shown in Figure \ref{fig:dnp_exp}b, where the effects of the nuclear dipole-dipole couplings are apparent in the doublet spectrum of the 100\% \cthirt diamond.


\begin{figure*}[h]
 \includegraphics[width=\textwidth]{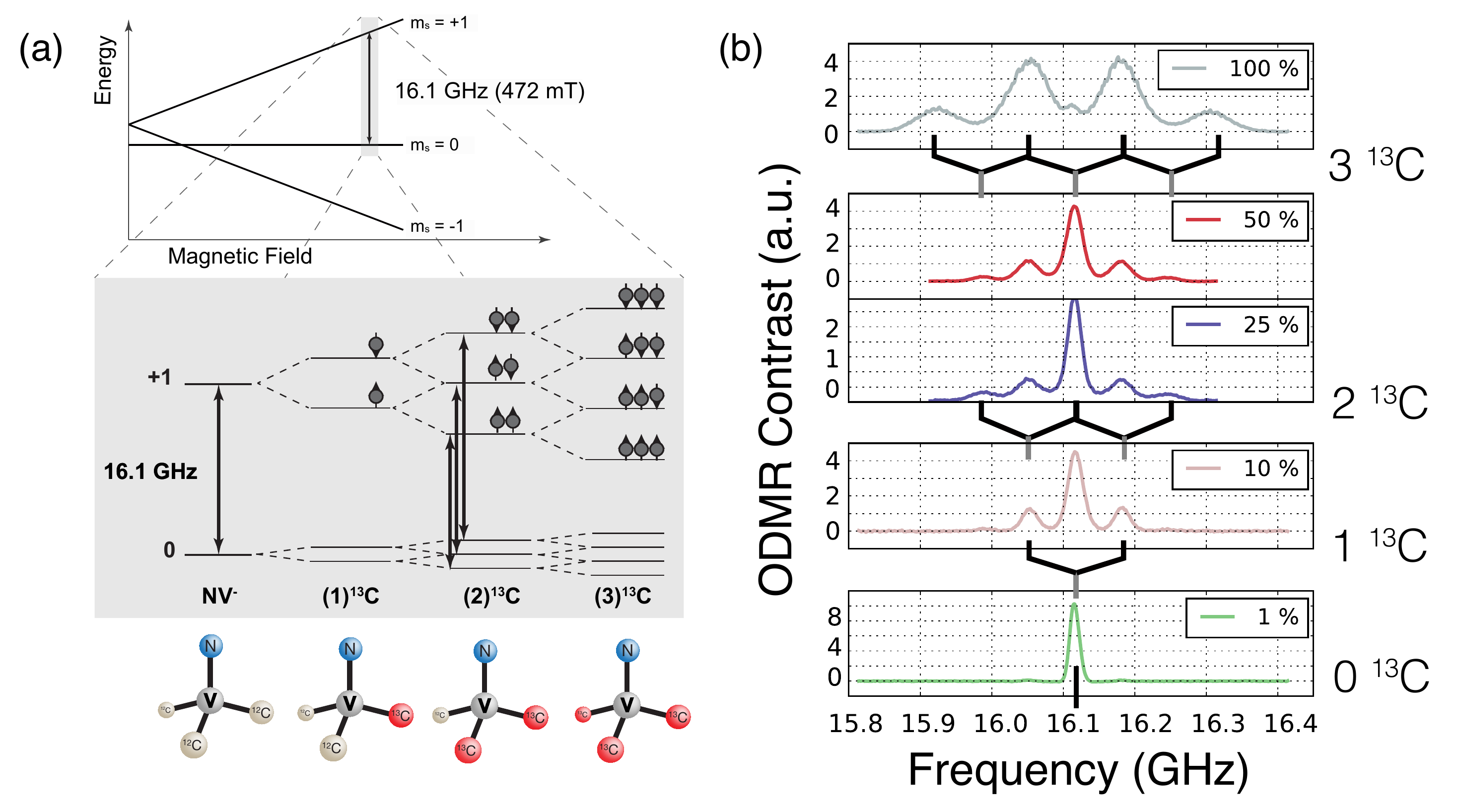}
 \caption{\label{fig:nvintro}  The effect of \cthirt enrichment on the energy level structure and ODMR spectroscopy of \nv centers. The energy level structure of the \nv as a function of magnetic field strength is shown in (a), where the magnetic field regime relevant to DNP experiments is highlighted and expanded. In this work, we focus on the high energy transition from the \nv \msz to \msp at 472 mT, which has a frequency of approximately 16.1 GHz. The strongest hyperfine interactions occur between the \nv and its first-shell \cthirt nuclei, those directly adjacent to it. Occupation of the first-shell sites leads to a splitting in not only the \nv \mspm levels, but also the \msz due to the anisotropic component of the coupling. The magnified energy level diagram shows how the \nv spin states split into hyperfine states determined by the different possible combinations of nuclear spin, as 1, 2, and 3 carbons are added. Diagrams of the \nv as well as the occupation of the first shell sites with \cthirt nuclei are given below the energy level diagram. The branching in these energy levels corresponds directly to the structure seen in the ODMR spectra of diamonds with varying \cthirt enrichment (b). A single line is observed for the \msz to \msp transition in \nv spin state for a sample with a natural abundance (1.1\%) of \cthirt, whereas a sample with 100\% \cthirt enrichment exhibits a quartet for the same transition.  \cite{Parker2015} }
 \end{figure*}

In a field of 472 mT, \nv spin transitions are observed at approximately 16.1 GHz and 10.3 GHz (Fig. \ref{fig:nvintro}a). In this work we focus on the higher-frequency transition. We read out the \nv spin state via optically-detected magnetic resonance (ODMR), where the \nv fluorescence intensity is monitored while sweeping the microwave frequency. Because the \mspm states are more likely to relax via intersystem crossing to the ground electronic state, resonant microwave excitation produces a detectable change in the \nv fluorescence intensity \cite{Doherty2013}. $^{13}$C spins within the first shell of nuclei, those directly adjacent to the \nv defect, are strongly coupled to the NV$^-$ spin and give a resolved hyperfine structure (Fig. \ref{fig:nvintro}). In a $^{13}$C enriched sample, the three nearest-neighbor sites are occupied by 0, 1, 2, or 3 $^{13}$C nuclei with the balance occupied by spinless $^{12}$C. The ODMR spectrum is therefore a superposition of 4 distinct patterns of hyperfine splittings with relative intensities determined by the degree of enrichment. The pattern of hyperfine splittings in Fig. \ref{fig:nvintro} is consistent with the known \cthirt-\nv hyperfine tensor for first-shell nuclear spins \cite{shim13}.

\begin{figure*}[h]
 \includegraphics[width=0.9\textwidth]{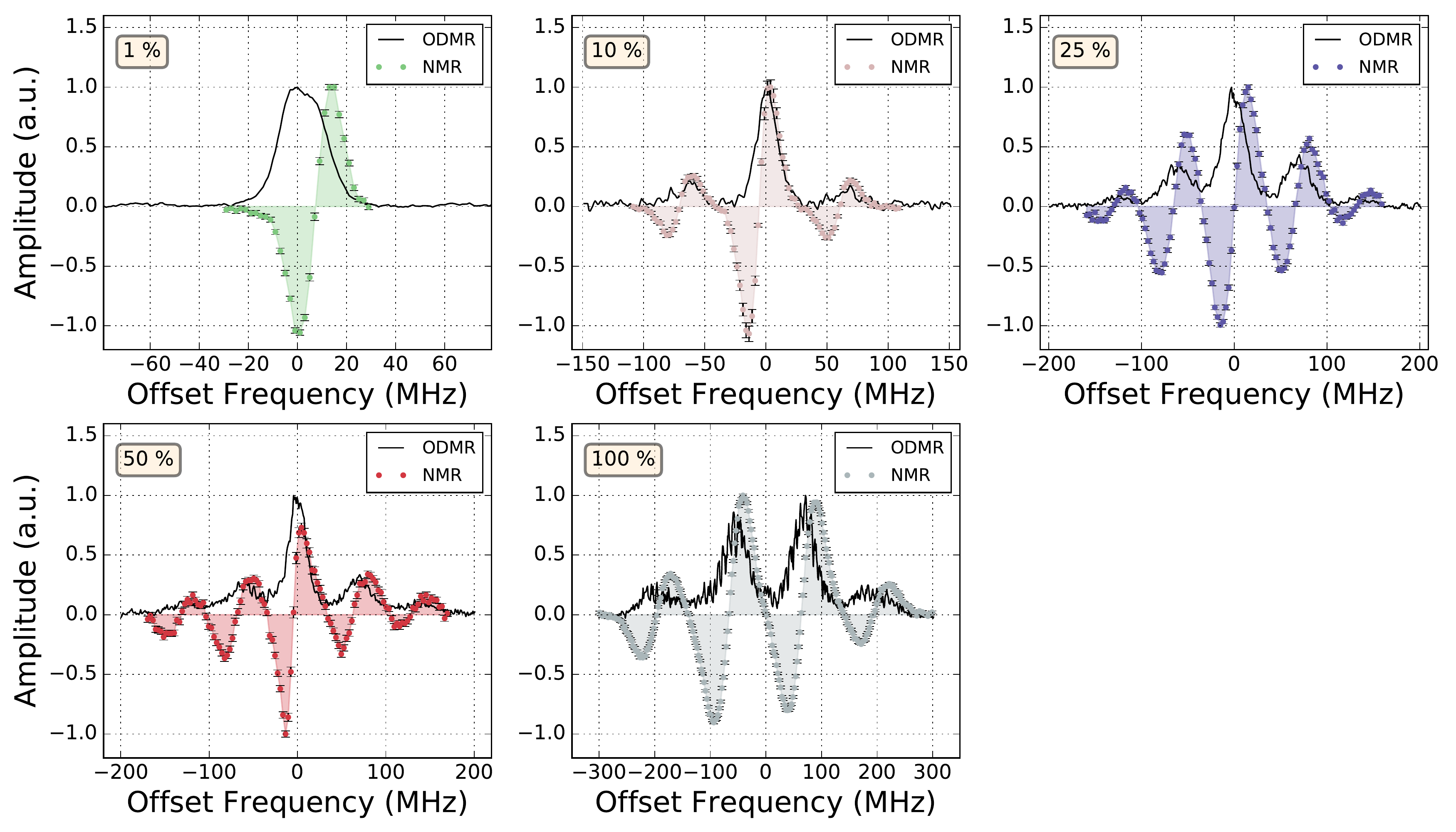}%
 \caption{\label{fig:se_curve} DNP spectra of the various diamond samples. Normalized hyperpolarized \cthirt NMR signal as a function of offset microwave frequency for each sample. The normalized ODMR data for each sample is given to show the corresponding high frequency transition of the \nv ESR spectrum. CW-DNP experiments are performed at 472.1 - 473.0 mT, thus the microwave frequency is centered at approximately 16.1 GHz for the various samples with an \cthirt NMR frequency of approximately 5.06 MHz. It should be noted the SNR of the ODMR spectra shown here differs from that of the ODMR spectra in Figure \ref{fig:nvintro} because the two were acquired at different microwave amplitudes. The optimum microwave amplitude for DNP measurements is higher than the microwave amplitude for optimized ODMR contrast (see Materials and Methods). }
 \end{figure*}

The DNP process involves driving NV$^-$ spin transitions to transfer polarization to nearby $^{13}$C spins. The microscopic mechanism of DNP depends on the width of the electron spin spectrum relative to the nuclear Larmor frequency and also on interactions between the electron spins\cite{DNPbook}. In our experiments the solid effect, cross effect, and thermal mixing mechanisms of DNP all potentially contribute, and are difficult to distinguish. Regardless of the local DNP mechanism, nuclear spin diffusion transports polarization to bulk $^{13}$C spins that do not interact directly with the \nv center. In general, the DNP intensity has an antisymmetric frequency dependence and is related to the intensity of the ODMR spectrum (Fig. \ref{fig:se_curve}), consistent with DNP mechanisms where the EPR spectrum is broader than the nuclear Larmor frequency. Interestingly, this trend holds for the satellite peaks induced by the strongly hyperfine-coupled $^{13}$C spins in the first shell, indicating that direct driving of the strongly-coupled hyperfine transition does not generate bulk polarization. This contrasts with \nv hyperpolarization near a level anti-crossing where highly mixed electron-nuclear spin states result in hyperpolarization of first-shell \cthirt which is then transported to bulk nuclear spins\cite{Alvarez2015}. Here, the symmetric intensities of the hyperfine-induced DNP satellites and the derivative DNP patterns at each satellite transition indicate the polarization of the first-shell $^{13}$C spins does not play a role in the bulk DNP, other than to induce a splitting of the NV$^-$ spectrum.



\begin{figure*}[!h]
 \includegraphics[width=0.4\textwidth]{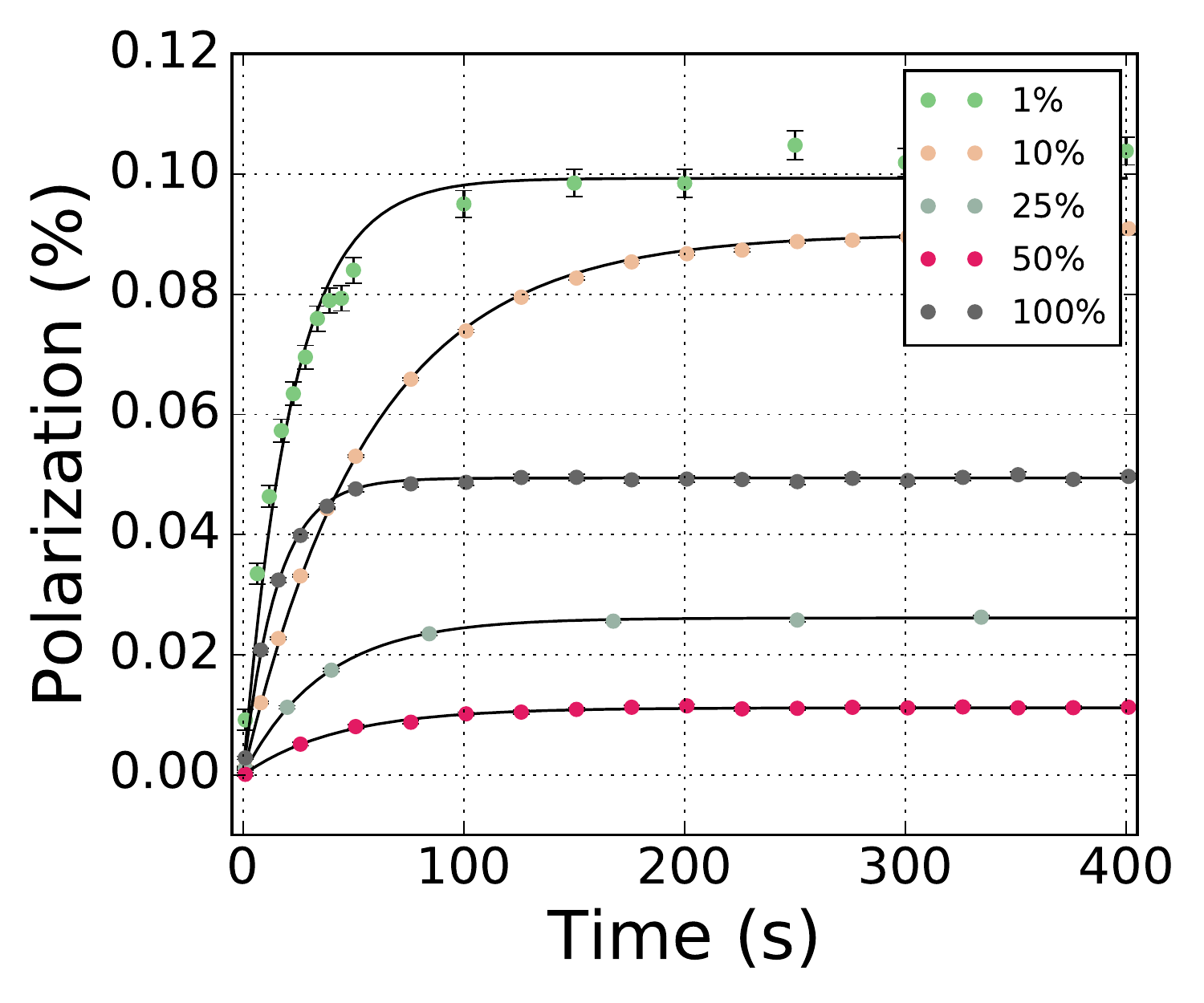}%
 \caption{\label{fig:buildup_pol} \cthirt polarization buildup curves are summarized for all samples involved in the study. Characteristic polarization buildup times \tdnp are given in Table \ref{tab:pol_results}. The sample with 100\% \cthirt enrichment gives the fastest buildup and the largest signal. The maximum polarization as well as the second-fastest buildup rate is observed from the sample with a natural abundance of \cthirt, which is likely due to it having the highest concentration of \nv centers (see Supplemental Information). }
 \end{figure*}

The level of maximum hyperpolarization is sample dependent due to the varied concentration of \cthirt, \nv centers, and other paramagnetic defects that cause nuclear spin relaxation, see Table \ref{tab:pol_results}. The highest level of polarization was achieved in the natural abundance diamond, which has the highest \nv concentration while the highest total magnetization occurred in the 100\% diamond. The 100\% \cthirt diamond also exhibited the fastest buildup of hyperpolarization (Fig. \ref{fig:buildup_pol}), suggesting that the enhanced rate of nuclear spin diffusion more efficiently transports polarization to bulk nuclei. All other parameters being equal, we expect 100\% \cthirt diamonds to be optimum for MRI contrast and polarization transfer applications. We have shown that, despite the spectral complexity associated with multiple strong hyperfine couplings, hyperpolarization can be efficiently transferred to bulk nuclear spins. Furthermore, the control we demonstrate over bulk nuclear spin polarization in samples with a high nuclear spin concentration provide insight for and enable development of quantum technologies employing strongly-coupled spin systems.

\section*{Acknowledgement}
This work was supported by the U.S. Department of Energy, Office of Science, Basic
Energy Sciences under Contract No. DE-AC02-05CH11231. The authors thank Dr. Ashok Ajoy for helpful discussions and Dr. Melanie Drake and Prof. Jeffrey Reimer for providing the natural abundance sample. This study was made possible by the help of Joseph Tabeling at Applied Diamond Inc./Delaware Diamond Knives for the custom synthesis of the samples used in this study.


\begin{table*}
\centering
\begin{tabular}{l|c|c|c|c|c} 
 Sample & D1 & D2 & D3 & D4 & D5 \\ [0.5ex] 
 \hline
 [\cthirt] (\%) & 1 & 10 & 25 & 50 & 100 \\
 \tdnp (s) & 22.34 $\pm$ 0.06 & 59.55 $\pm$ 0.03 & 36.14 $\pm$ 0.02 & 42.94 $\pm$ 0.04 & 15.28 $\pm$ 0.02 \\
 Enhancement & 1264 $\in$ [854,2430] & 1094 $\pm$ 202 & 318 $\pm$ 22 & 138 $\pm$ 4 & 604 $\pm$ 11 \\
 P$_{enh}$ (\%) & 0.10 $\in$ [0.071,0.20] & 0.091 $\pm$ 0.017 & 0.026 $\pm$ 0.002 & 0.011 $\pm$ 0.0004 & 0.050 $\pm$ 0.0009 \\
\end{tabular}
\caption{ Summary of the polarization buildup time \tdnp, enhanced \cthirt nuclear polarization (P$_{enh}$) and maximum DNP enhancement for each diamond sample. }
\label{tab:pol_results}
\end{table*}

\section{Materials and Methods}

$^{13}$C enriched diamonds were grown by chemical vapor deposition (Applied Diamond Inc.) using $^{13}$C enrichments of methane with 600 ppm nitrogen as a precursor. $^{13}$C concentrations of 10\%, 25\%, 50\%, and 100\% $^{13}$C were used. These samples were compared to a sample grown by high-pressure high-temperature (HPHT) diamond synthesis (Sumitomo Electric Carbide, Inc.) with a substitutional nitrogen concentration of approximately 200 ppm and \cthirt concentration of 1.1\% (natural abundance). All samples were irradiated with 1 MeV electrons at a fluence of $10^{18}$ cm$^{-2}$ (Prism Gem LLC) and annealed at 800\degree C for 2 h to produce an NV$^-$ concentration of 1-10 ppm. Optically detected magnetic resonance (ODMR) and DNP were performed in a purpose-built instrument consisting of a custom probe fixed in an electromagnet (GMW Associates, Model 3472-50 with Danfysik 858 power supply). The probe includes a radiofrequency circuit for inductive NMR excitation and detection, a goniometer for two-axis alignment of the defect axis along the magnetic field, a 3 mm wire loop for microwave excitation (Agilent E8257D signal generator), and optical access to the sample (532 nm Coherent Verdi G5 laser). For all DNP experiments, the amplitude of the microwave frequency was set to 10 dBm and amplified with 3 W amplifier (Mini-Circuits ZVE-3W-183+) before being sent to the 3 mm loop. The NMR circuit includes a 30-turn planar coil of 46 AWG copper wire, with capacitance added to impedance-match the circuit at 5.06 MHz. The NMR component of DNP experiments are carried out with a Magritek Kea$^2$ console. For ODMR measurements, the microwave amplitude was modulated 100\% at 200 Hz from the reference signal of the lock-in amplifier (Stanford Research Systems, SR830) and the fluorescence signal was detected with an avalanched photodiode (APD 410A, Thorlabs). The lock-in amplifier measured the in-phase component of the fluorescence signal at the modulation frequency using a time constant of 30 ms. A spectrum was acquired by stepping through a range of microwave frequency centered on the \nv ESR, where each step consists of changing the microwave frequency by one step, waiting 50 ms, and measuring the ODMR signal from the lock-in. These spectra were used to characterize the samples, align the defect axis along the magnetic field, and set the strength of the magnetic field as measured by the ensemble of \nv defects for DNP experiments. 

DNP experiments were carried out at 472.2 mT. A schematic of the DNP experiment is shown in Figure \ref{fig:dnp_exp}. A 532 nm laser with a beam diameter of approximately 5 mm is set to an output power of 1 W/cm$^2$ and applied continuously throughout experiments. The laser beam is kept large to irradiate the full surface of the diamond. A set of 90\degree pulses are used to destroy thermal \cthirt polarization before waiting a time \trec for \cthirt polarization to build as a result of DNP processes. The microwave frequency is set to \fsat and applied continuously for the duration of \trec. A simple 90 pulse-acquire experiment is then used to determine the \cthirt NMR signal. This is repeated for a range of \fsat centered on the \nv ESR to acquire the DNP spectrum of \cthirt NMR signal as a function of microwave frequency \fsat. The \cthirt NMR signal was compared by fitting the free induction decay (FID) at $f_{MW,i}$ to the FID of the signal with maximum enhancement by a scaling factor. All data are reported with error bars giving 95\% confidence intervals for the scaling factors, taken from the standard deviation of the parameter estimates of the fit. All NMR raw data were acquired with Prospa (software supplied with the Kea$^2$ spectrometer) and exported for processing in Python with Matplotlib \cite{Hunter2007}, SciPy and NumPy \cite{scipynumpy, scipylib}  packages. Additional details on methods and data analysis can be found in the Supplemental Information.

Photoluminescence (PL) experiments in the Supplemental Information document were carried out with a homebuilt confocal microscope to gain a qualitative understanding of the defect content in each of the samples. The confocal microscope involves a 532 nm laser (Opto Engine LLC, MGL-III-532-200mW) directed to the sample through a microscope objective with NA = 0.4 (Nikon M Plan 20 ELWD). From this NA we estimate an excitation volume of $\SI{1.9e5}{\mu m^3}$. The objective is also used to collect the fluorescence and direct it through a dichroic mirror to a spectrometer (Mightex HRS-BD1-025). The emission of the samples were collected using an approximate optical excitation power of $\SI{2.5}{kW/cm^2}$. We use the minimum optical power required to detect the emission spectra with the Mightex Spectrometer in order to suppress any changes in the emission spectra due to charge-state conversion between \nv and \nvn. Emission spectra were collected from sixteen random points in each of the samples. Each spectrum was averaged 256 times with an exposure time of 100 ms. The raw data were acquired using the software provided by Mightex for interfacing with the spectrometer, and exported for processing in Python.  
\bibliography{dnp_13c}

\clearpage

\section{Supplemental Information}

In this document we summarize all data supporting the analysis of results and conclusions drawn in the body of the manuscript. All confidence intervals reported in this work correspond to 95\% bounds. 

\begin{table*}
\centering
\begin{tabular}{l|c|c|c|c|c} 
 Sample & D1 & D2 & D3  & D4 & D5 \\ [0.5ex] 
 \hline
 [\cthirt] (\%) & 1 & 10 & 25 & 50 & 100 \\
 Mass (mg) & 4.4 & 12.4 & 10.6 & 5.9 & 9.3  \\
 \tdnp (s) & 22.34 $\pm$ 0.06 & 59.55 $\pm$ 0.03 & 36.14 $\pm$ 0.02 & 42.94 $\pm$ 0.04 & 15.28 $\pm$ 0.02 \\
 \tonecth (s) & 13.08 $\pm$ 1.11 & 76.85 $\pm$ 3.70 & 51.18 $\pm$ 0.42 & 16.68 $\pm$ 2.20 & 10.33 $\pm$ 0.14 \\
 Enhancement & 1264 $\in$ [854,2430] & 1094 $\pm$ 202 & 318 $\pm$ 22 & 138 $\pm$ 4 & 604 $\pm$ 11 \\
 P$_{enh}$ (\%) & 0.10 $\in$ [0.071,0.20] & 0.091 $\pm$ 0.017 & 0.026 $\pm$ 0.002 & 0.011 $\pm$ 0.0004 & 0.050 $\pm$ 0.0009 \\

\end{tabular}
\caption{ Summary of parameters and maximum DNP enhancement for diamond samples. Values for \tdnp, \tonecth, the enhancement from DNP processes, and the enhanced \cthirt nuclear polarization (P$_{enh}$) are measured at 470 mT.}
\label{tab:char}
\end{table*}

\subsection{Sample characterization}

\begin{figure*}[h!]
\centering
\includegraphics[width=0.5\textwidth]{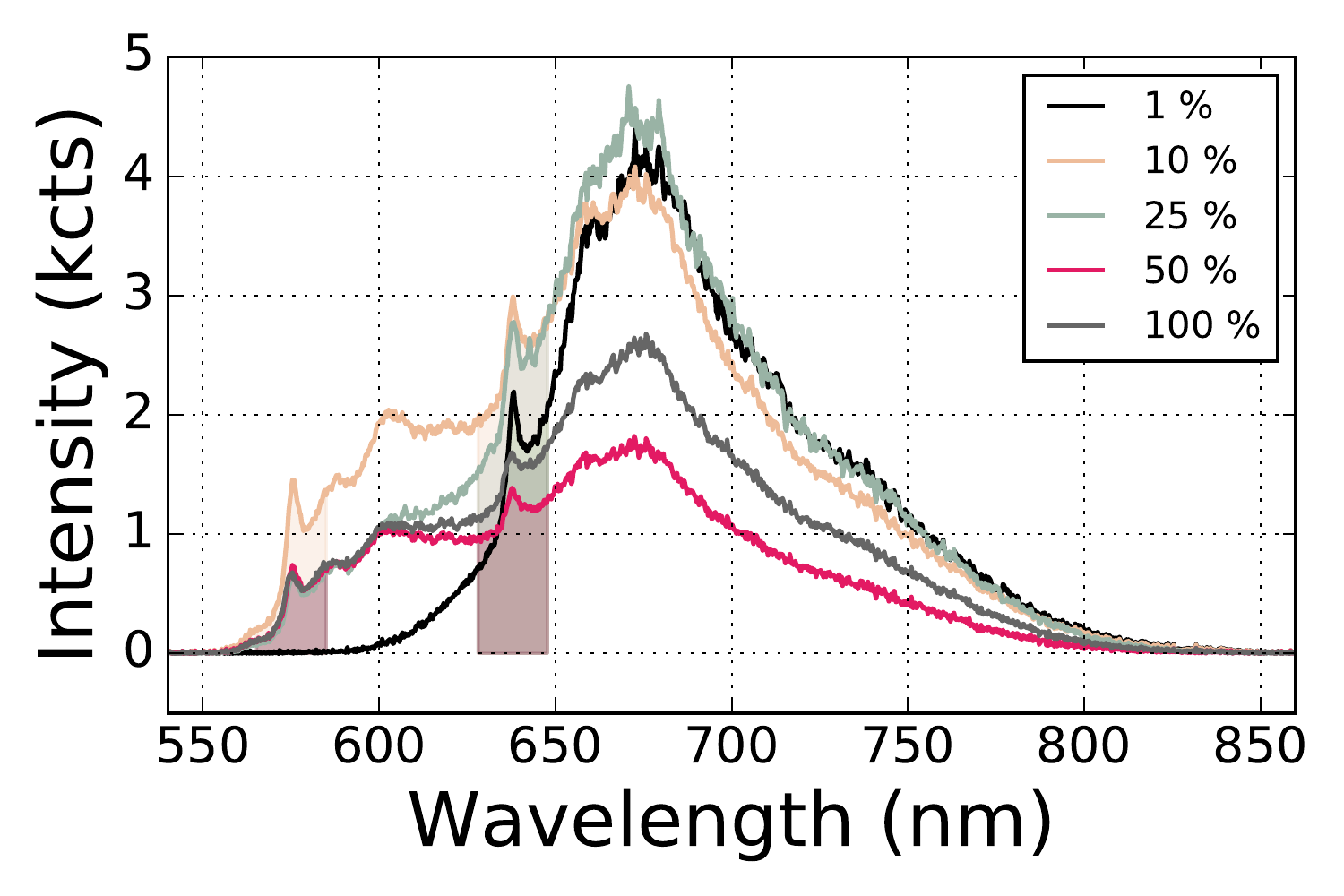}%
\caption{\label{fig:pldata} Photoluminescence data. The trace for the emission from D1 (1\% \cthirt) is given in black to emphasize the difference in emission between HPHT-grown and CVD-grown samples. The majority of NV centers in sample D1 are negatively-charged. The \nvn and \nv ZPLs are highlighted. }
\end{figure*}

The samples acquired for this study include one with a natural abundance of \cthirt grown by high-pressure high-temperature (HPHT) methods (D1) in addition to four samples grown by chemical vapor deposition (CVD, Applied Diamond Inc.). The CVD samples (D2-D5) were grown with with added nitrogen precursor and varying levels of \cthirt enriched methane to accomplish a distribution of \cthirt concentrations. Sample parameters are summarized in Table S\ref{tab:char}. The photoluminescence (PL) spectrum of each sample (Fig. S\ref{fig:pldata}) was acquired in order to order to gain a qualitative understanding of the sample composition. The details on how the PL spectra were acquired are given in the Materials and Methods section. The PL spectrum of a diamond containing a high concentration of \nv centers typically has four prominent features within the range of 560 - 750 nm. The two sharp peaks found roughly at 575 and 638 nm correspond to the zero phonon lines of the neutral (\nvn) and negatively-charged \nv, respectively. The broad features red-shifted from each ZPL are the phonon side bands corresponding to higher vibrational levels of the excited states of each defect. 

Most importantly, the PL spectra qualitatively demonstrate a large variation in the concentration of \nv, \nvn, and substitutional nitrogen (\pone, also called P1 centers) among even the CVD samples, preventing quantitative analysis of the polarization buildup rates. Also of note from the PL spectra is that sample D1 has a much larger ratio of \nv:\nvn than any of the CVD-grown samples. This is consistent with previous studies \cite{kennedy2003long} and is attributed to higher levels of \pone, typical of HPHT growth, which act as donors to the \nvn to achieve larger conversion to \nv. These data indicate a higher \nv concentration in sample D1, which provides an explanation for why the polarization buildup rate of sample D1 exceeds most of the buildup rates of the \cthirt-enriched samples despite slow spin diffusion. Polarization buildup (\tdnp) and \cthirt nuclear spin-lattice relaxation times (\tonecth) are summarized in Table S\ref{tab:char}.

\subsection{Calibration of the \cthirt nuclear polarization enhancement in diamond.}

\begin{figure*}[h!]
\centering
\includegraphics[width=0.4\textwidth]{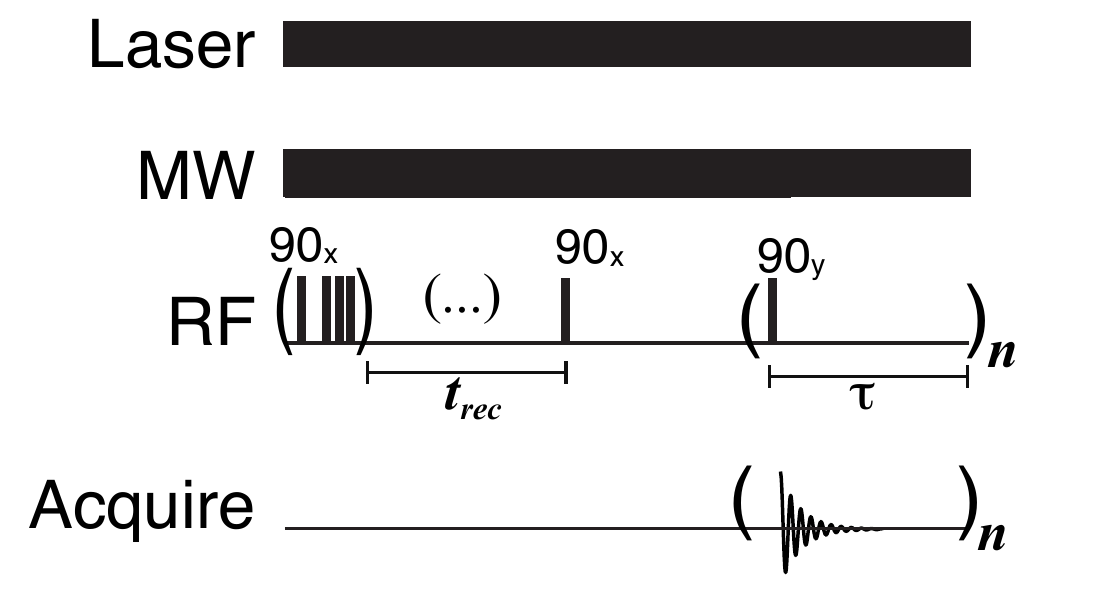}
\caption{\label{fig:sechoes} Schematic of the solid echo pulse sequence for detecting hyperpolarized \cthirt NMR at 5 MHz. The laser and microwave excitation are applied continuously, where the microwave frequency is set to the optimum frequency for DNP. The sequence begins with a series of 90\degree saturation pulses to ensure zero initial \cthirt polarization. Polarization builds up for a time of $t_{rec}$. The solid echo sequence is initiated with a 90\degree pulse along the x axis, a delay of time $\tau$ called the ``echo time", following by a train of $n$ 90\degree pulses along the y axis separated by $\tau$. A free-induction decay is acquired after every 90$_y$\degree pulse. The 90$_y$ pulses follow a phase cycle of [180, 0, 0, 180].}
\end{figure*}

The long nuclear spin lattice relaxation times of \cthirt in diamond as well as the low sensitivity of \cthirt NMR at 5 MHz make detecting a thermal \cthirt NMR signal challenging. We overcome this challenge by using a solid echo pulse sequence \cite{boden1973} (shown schematically in Figure \ref{fig:sechoes}) which extends the duration of the free induction decay via a spin-locking effect. The pulse sequence is originally carried out on the hyperpolarized \cthirt NMR signal, which is fit with a biexponential function. We assume the thermally polarized signal has the same functional form, and we seek the scaling factor that relates the thermal and hyperpolarized signals. The schematic in Figure \ref{fig:sechoes} shows detection of the hyperpolarized signal, which includes laser and microwave excitation at the optimized frequency. For measurement of the thermal signal, the same pulse sequence is used with the laser off, and the microwave frequency set to 13 GHz (far off resonance from any transition contributing to DNP). 

For solid echo pulse measurements to be successful, the NMR 90 pulse length should be short relative to the evolution of the spins and $\tau$ should be short as possible relative to the decay of the FID to efficiently spin-lock the nuclear polarization in the transverse plane, otherwise coherence is lost rapidly and the spin-locking effect is diminished. In our experiment, 90 pulse lengths were in the range 1.0 - $\SI{2.5}{\mu s}$. Typical parameters for an echo sequence with these samples involved acquiring the free-induction decay with 32 points at a dwell time of $\SI{0.5}{\mu s}$ and $\tau$ of $\SI{40}{\mu s}$ for 500 echoes. For the hyperpolarized signal, the polarization was allowed to build for a time of 4\tdnp (see Table S\ref{tab:char}).

\begin{figure*}[h!]
\includegraphics[width=0.9\textwidth]{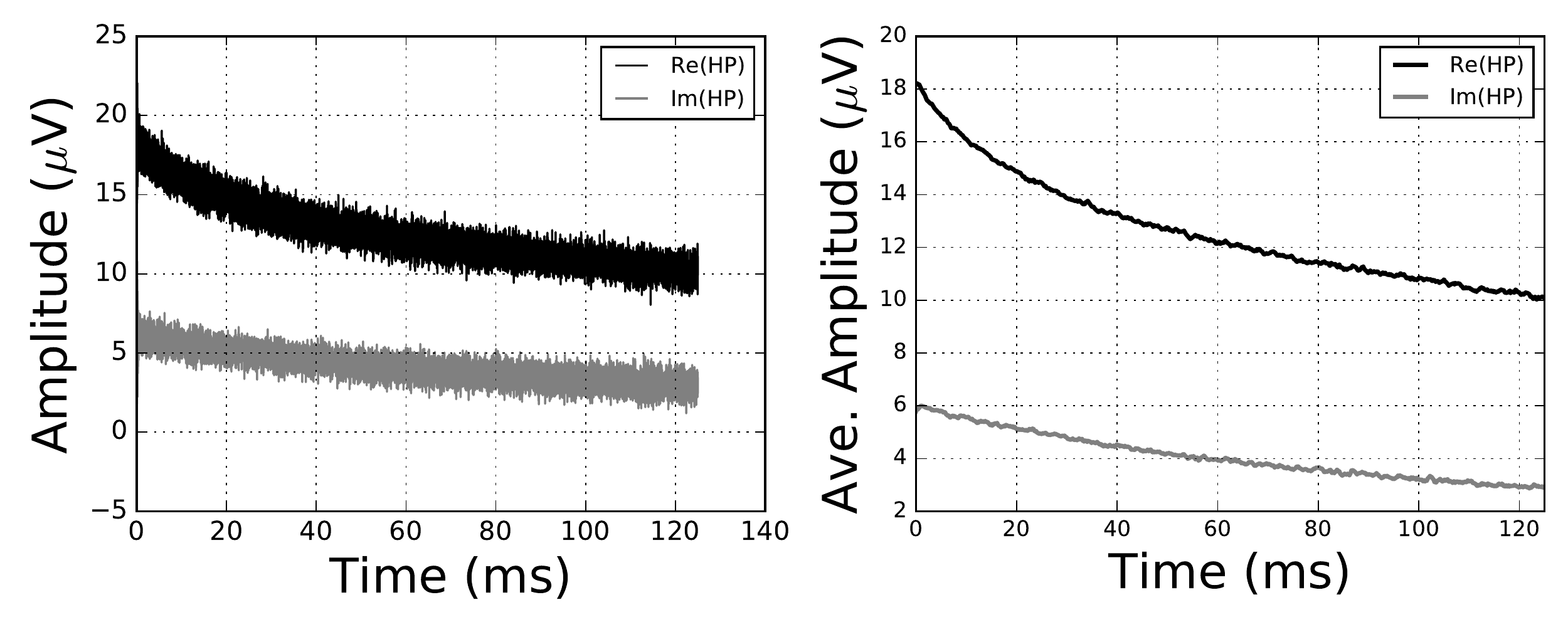}
\caption{\label{fig:mavg} Raw data from the solid echo pulse sequence is given on the left, while the data processed with a moving average is on the right.}
\end{figure*}

The solid echo pulse sequence results in a time domain decay extended well beyond T$_2^*$ of a true FID. We process the solid echo data with a moving average to smooth short timescale noise unrelated to the spin dynamics. The size of the window of the moving average varies based on the dynamics of the spin system, but ranges from 0.1 - 0.5 ms. Figure \ref{fig:mavg} demonstrates this processing. 

\begin{figure*}[h!]
\begin{subfigure}{0.48\textwidth}
\includegraphics[width=\textwidth]{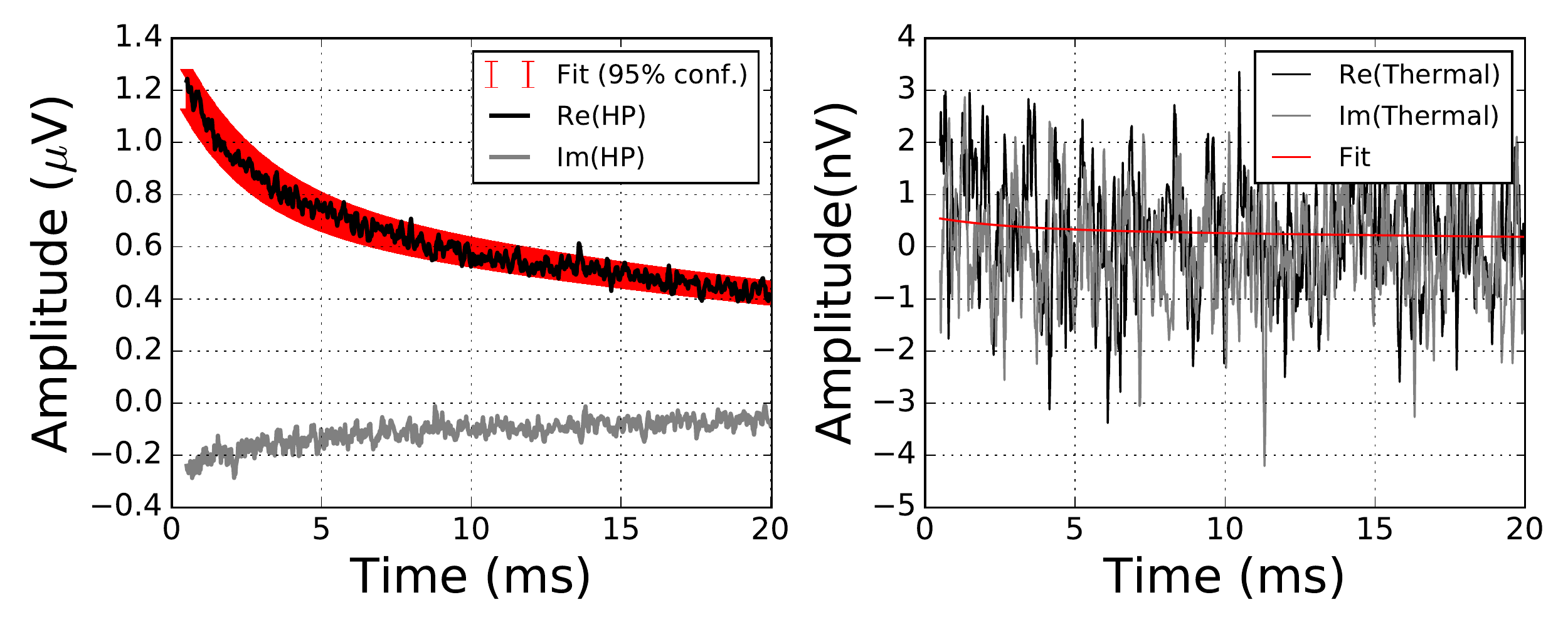}
\caption{D1: 1\% \cthirt. Fitted \ttwo values from the biexponential model of the hyperpolarized signal were T$_{2,1}$ = 2.43 $\pm$ 0.17 ms and T$_{2,2}$ = 33.17 $\pm$ 1.63 ms. The amplitude of the thermal signal fit by scaling to the hyperpolarized signal is corrected by a factor of 1.758.}
\end{subfigure}
\hfill
\begin{subfigure}{0.48\textwidth}
\includegraphics[width=\textwidth]{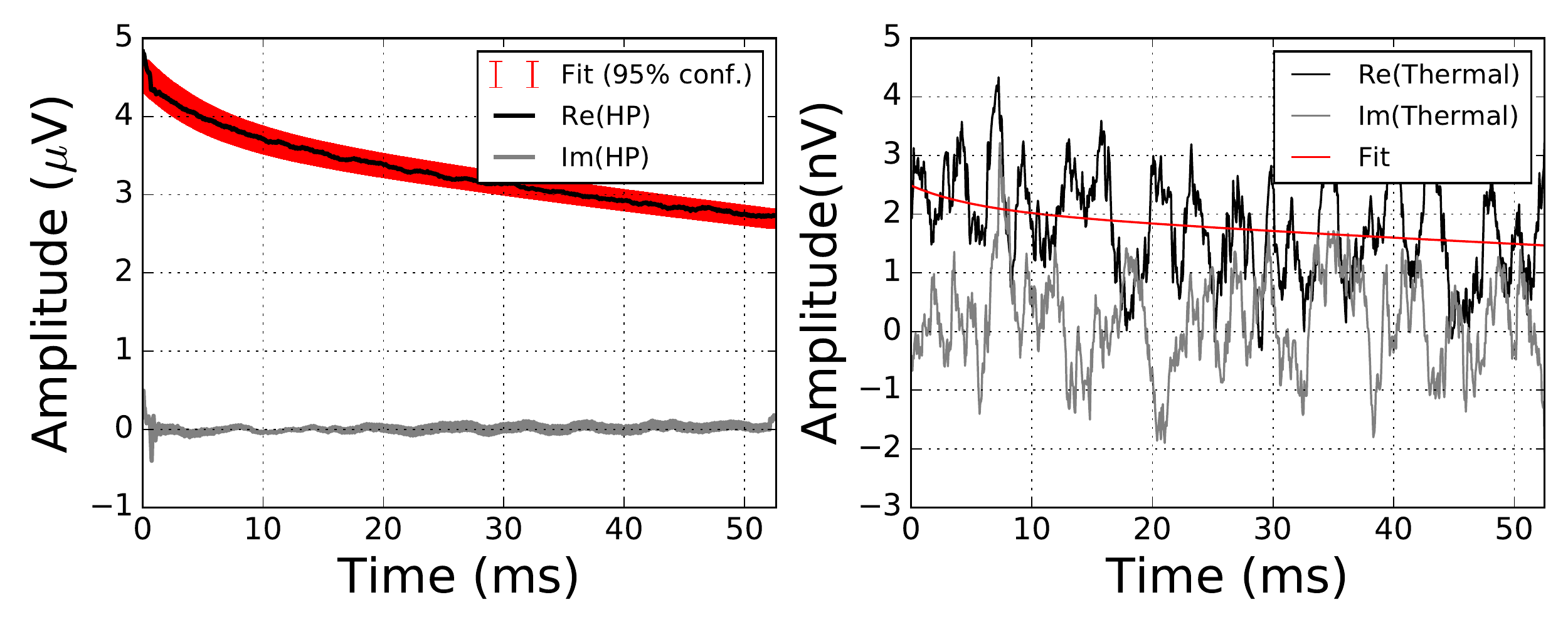}
\caption{D2: 10\% \cthirt. Fitted \ttwo values from the biexponential model of the hyperpolarized signal were T$_{2,1}$ = 5.22 $\pm$ 0.22 ms and T$_{2,2}$ = 146.67 $\pm$ 1.78 ms. The amplitude of the thermal signal fit by scaling to the hyperpolarized signal is corrected by a factor of 1.673.}
\end{subfigure}
\vfill
\begin{subfigure}{0.48\textwidth}
\includegraphics[width=\textwidth]{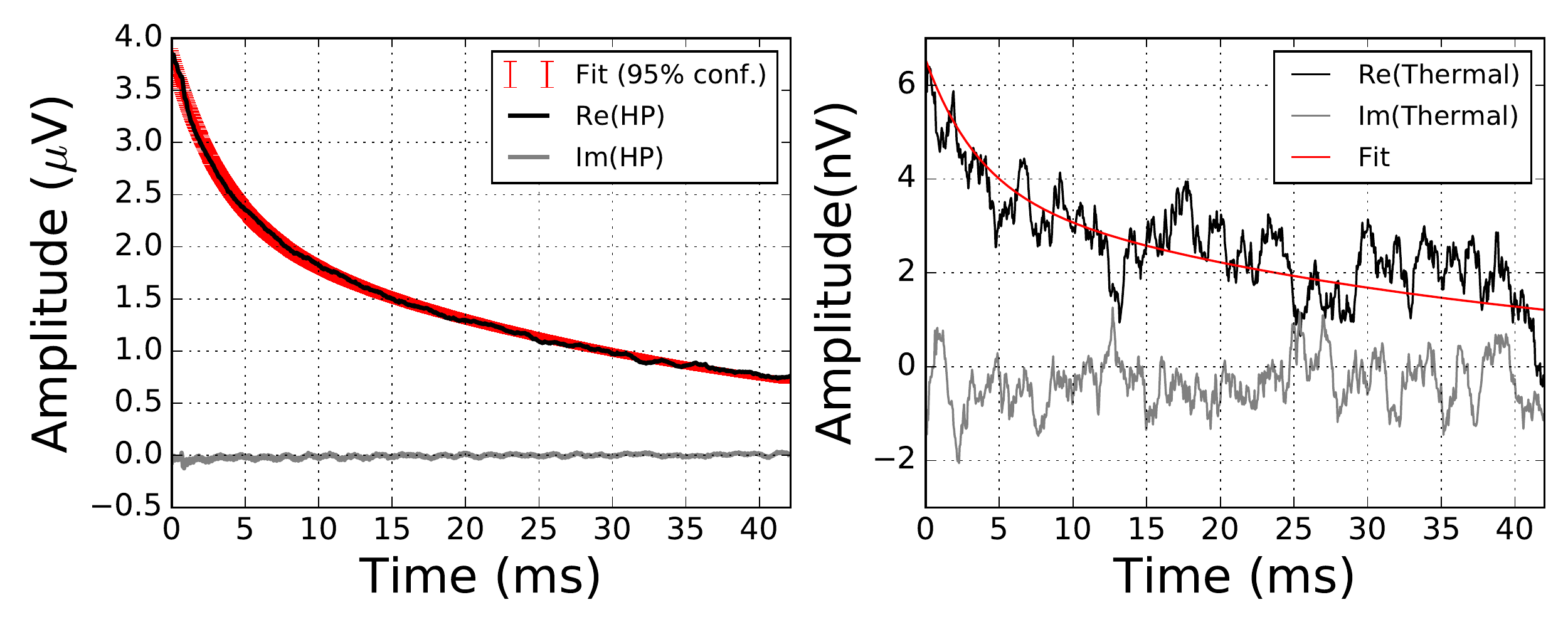}
\caption{D3: 25\% \cthirt. Fitted \ttwo values from the biexponential model of the hyperpolarized signal were T$_{2,1}$ = 3.56 $\pm$ 0.06 ms and T$_{2,2}$ = 36.54 $\pm$ 0.29 ms. The amplitude of the thermal signal fit by scaling to the hyperpolarized signal is corrected by a factor of 1.844.}
\end{subfigure}
\hfill
\begin{subfigure}{0.48\textwidth}
\includegraphics[width=\textwidth]{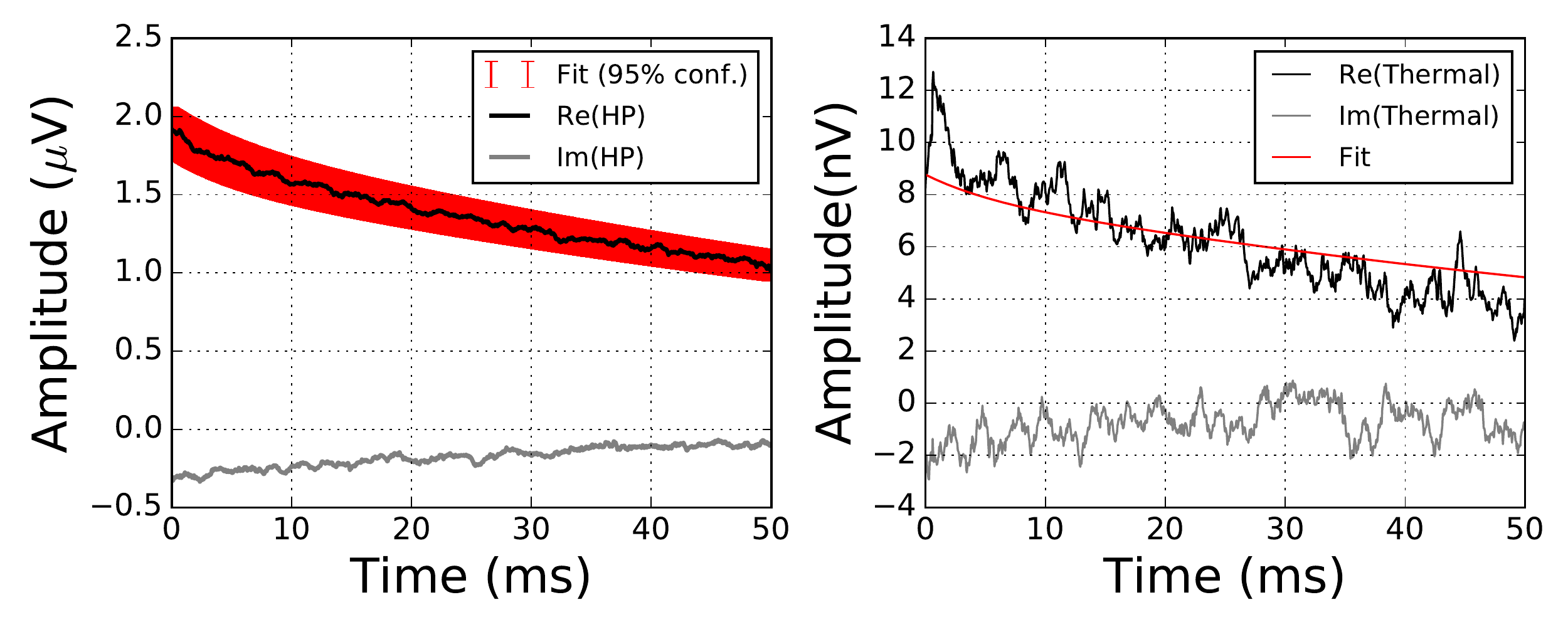}
\caption{D4: 50\% \cthirt. Fitted \ttwo values from the biexponential model of the hyperpolarized signal were T$_{2,1}$ = 5.42 $\pm$ 0.51 ms and T$_{2,2}$ = 100.36 $\pm$ 1.22 ms. The amplitude of the thermal signal fit by scaling to the hyperpolarized signal is corrected by a factor of 1.565.}
\end{subfigure}
\vfill
\begin{subfigure}{0.48\textwidth}
\includegraphics[width=\textwidth]{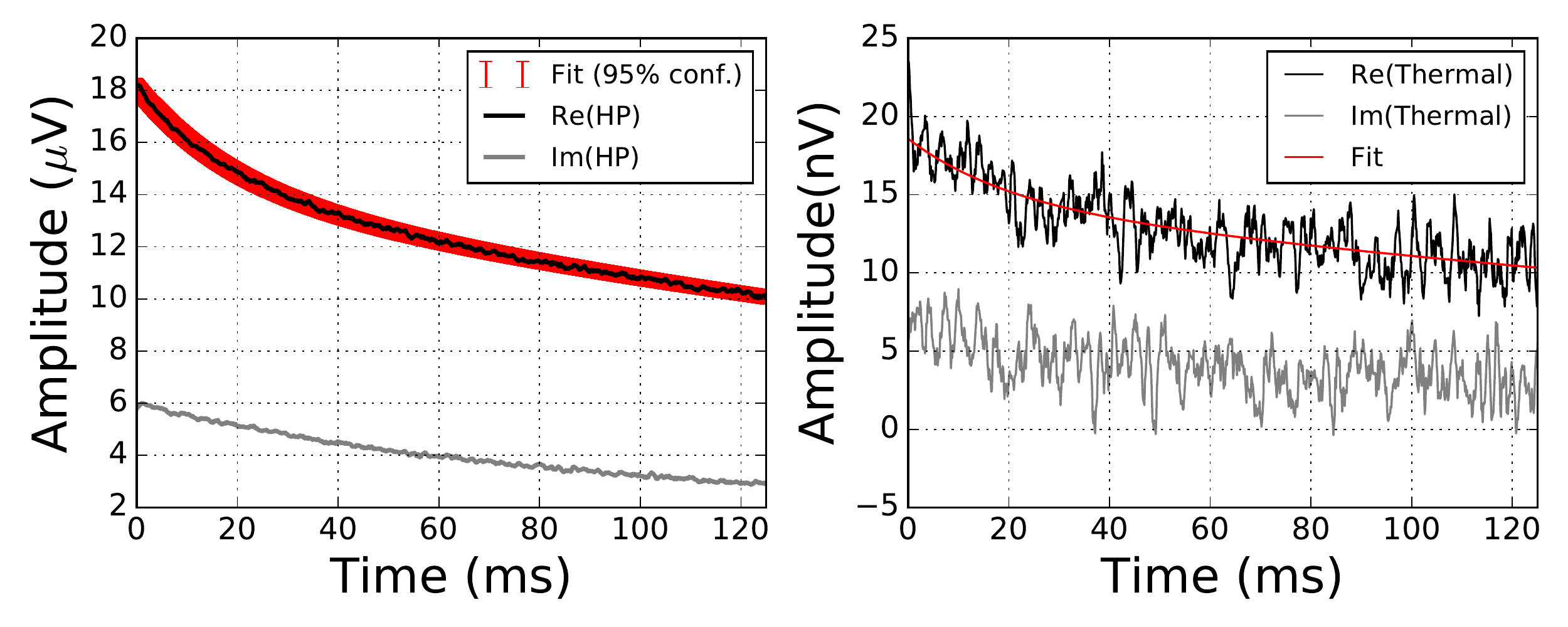}
\caption{D5: 100\% \cthirt. Fitted \ttwo values from the biexponential model of the hyperpolarized signal were T$_{2,1}$ = 19.48 $\pm$ 0.36 ms and T$_{2,2}$ = 364.47 $\pm$ 3.97 ms. The amplitude of the thermal signal fit by scaling to the hyperpolarized signal is corrected by a factor of 1.612.}
\end{subfigure}
\hfill
\caption{\label{fig:allt1} \ttwo measurements of all samples, showing hyperpolarized \cthirt NMR signal on the left, thermal \cthirt NMR signal on the right.  Note the change of scale in each figure; the amplitude for each hyperpolarized signal is in $\SI{}{\mu V}$ while the amplitude for each thermal signal is in nV. }
\end{figure*}



\begin{figure*}[h!]
\begin{subfigure}{0.3\textwidth}
\includegraphics[width=\textwidth]{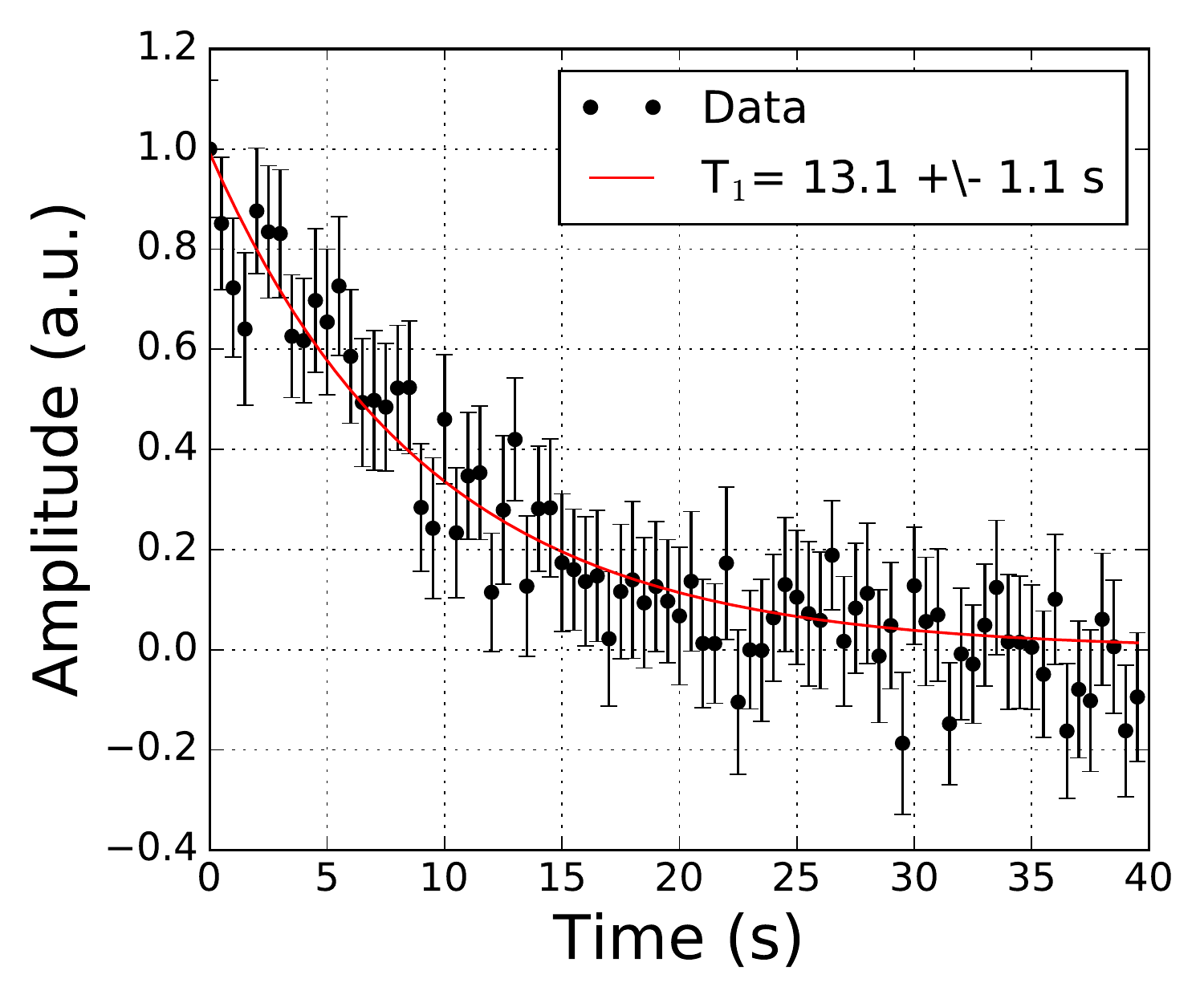}
\caption{D1: 1\% \cthirt. Small-flip pulse angle of 10.12\degree used.}
\end{subfigure}
\hfill
\begin{subfigure}{0.3\textwidth}
\includegraphics[width=\textwidth]{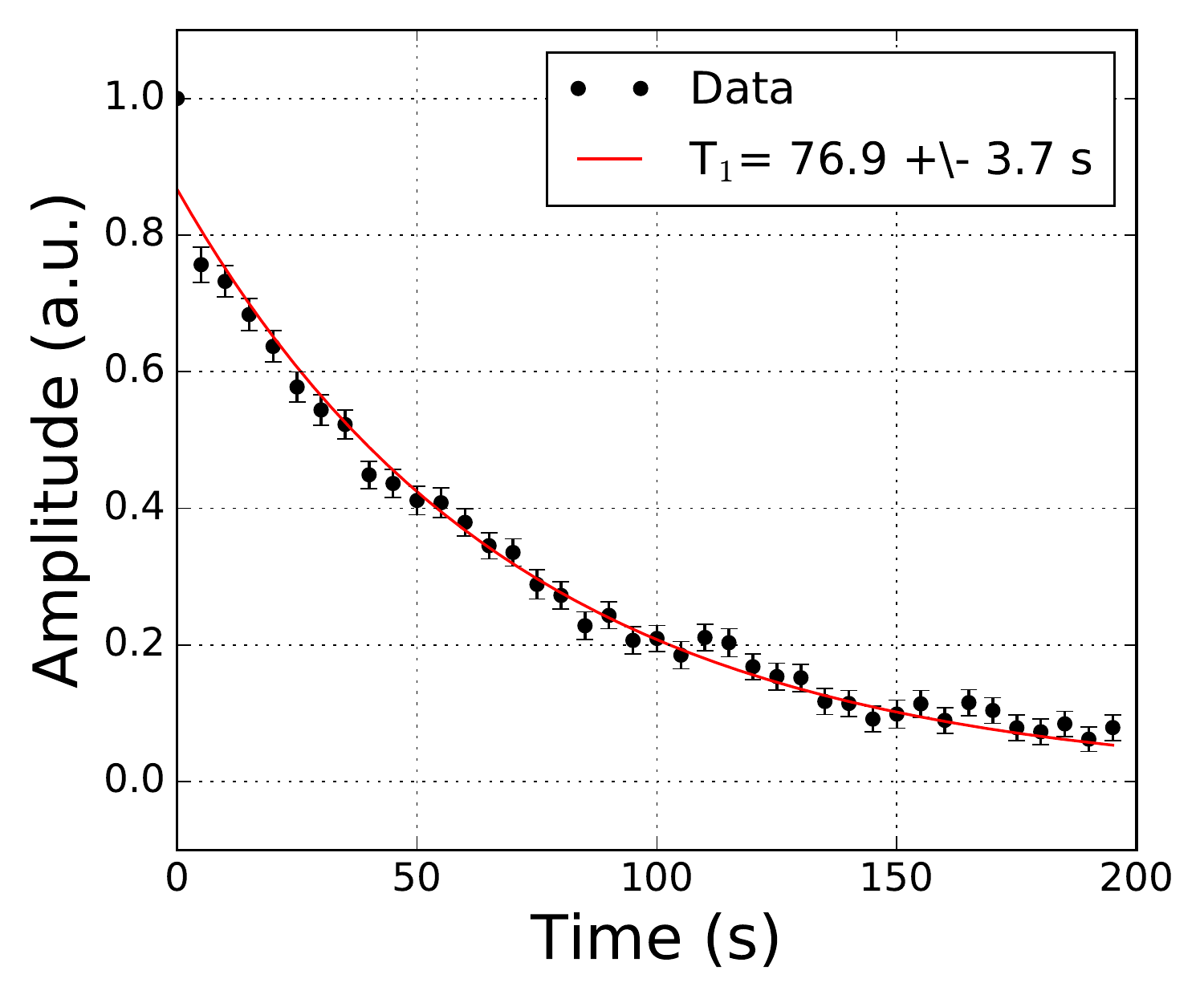}
\caption{D2: 10\% \cthirt. Small-flip pulse angle of 6.44\degree used.}
\end{subfigure}
\hfill
\begin{subfigure}{0.3\textwidth}
\includegraphics[width=\textwidth]{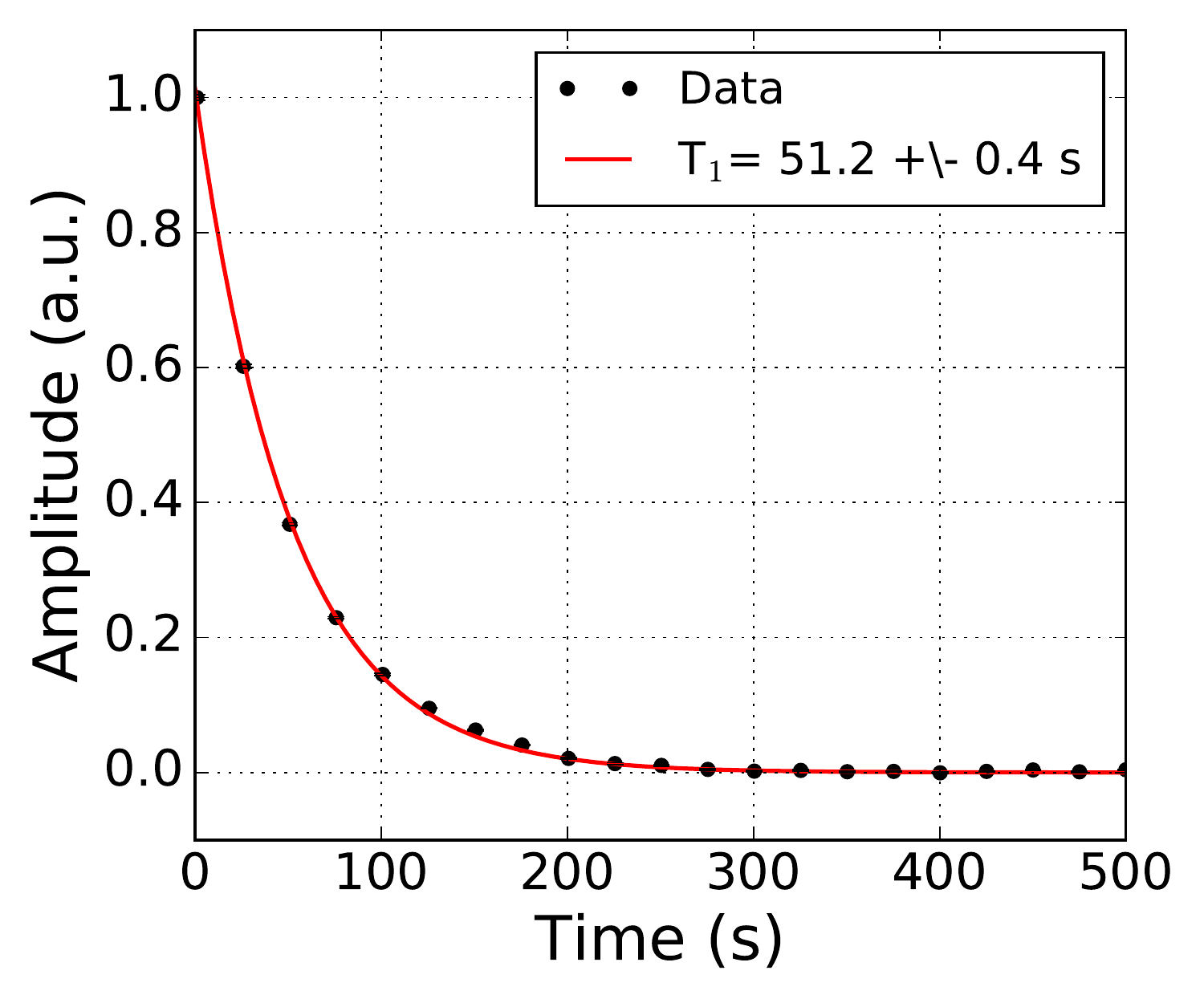}
\caption{D3: 25\% \cthirt. \tone data here collected point-by-point.}
\end{subfigure}
\begin{subfigure}{0.3\textwidth}
\includegraphics[width=\textwidth]{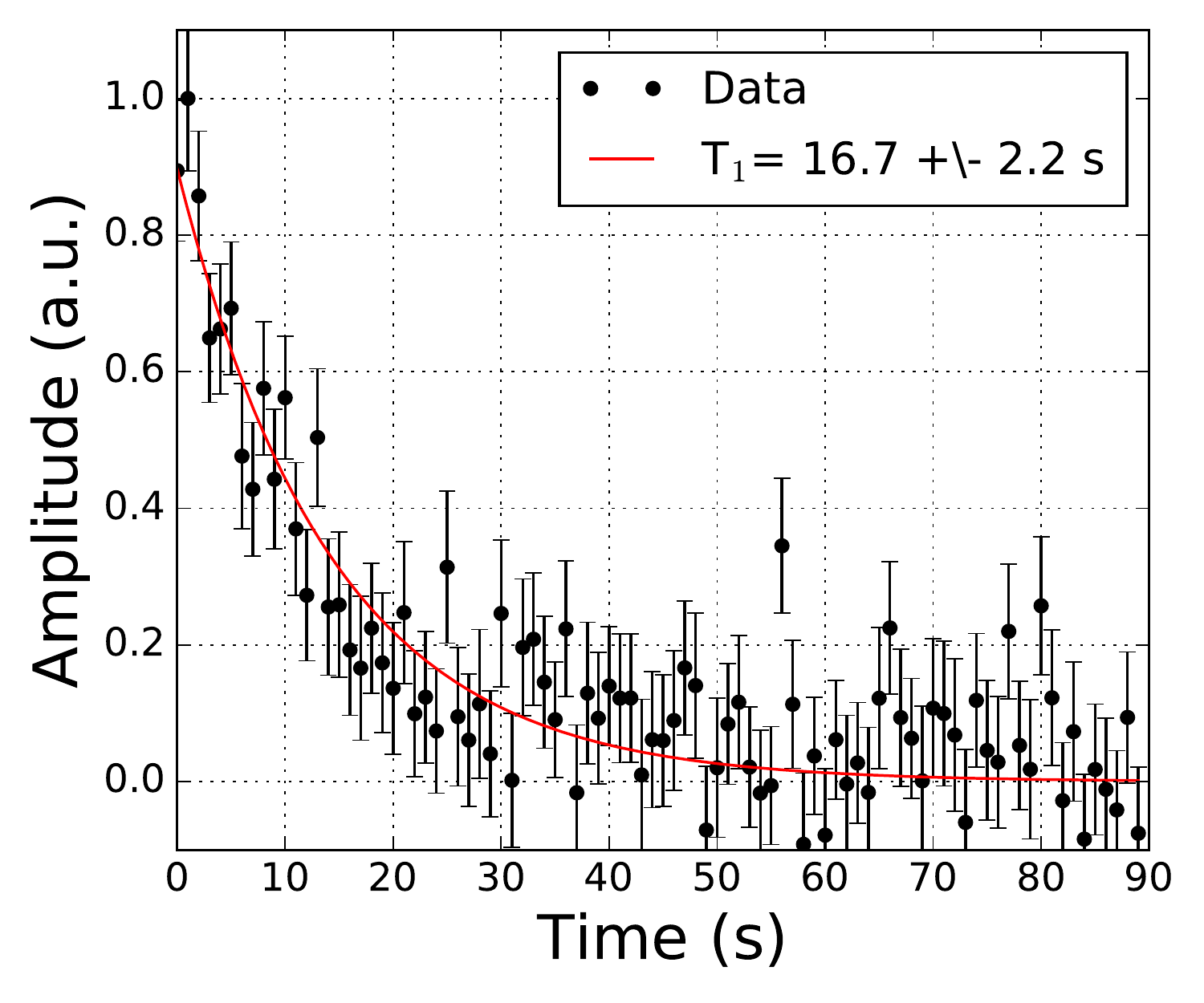}
\caption{D4: 50\% \cthirt. Small-flip pulse angle of 8.27\degree used.}
\end{subfigure}
\hfill
\begin{subfigure}{0.3\textwidth}
\includegraphics[width=\textwidth]{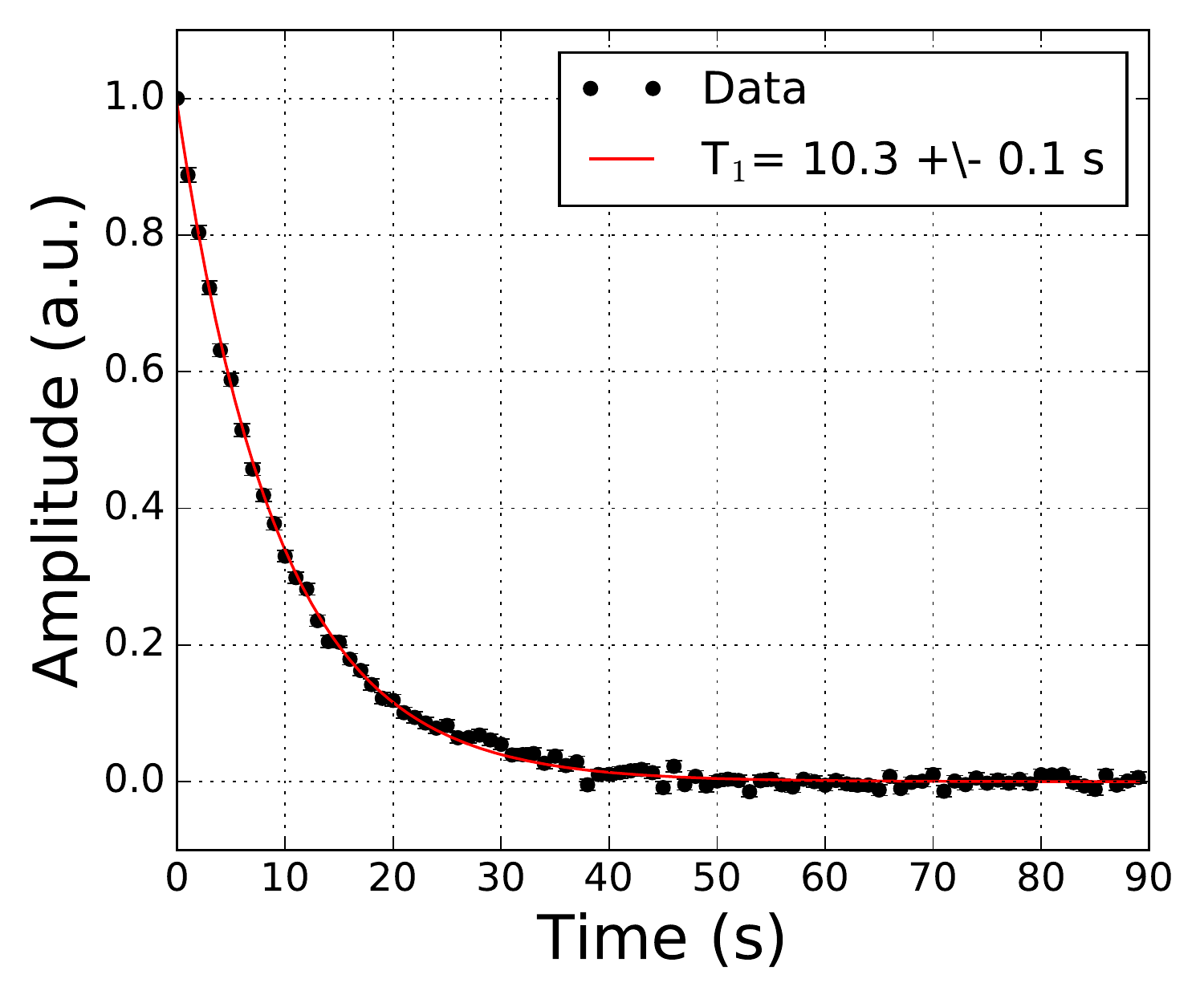}
\caption{D5: 100\% \cthirt. Small-flip pulse angle of 8.27\degree used.}
\end{subfigure}
\hfill
\begin{subfigure}{0.3\textwidth}
\includegraphics[width=\textwidth]{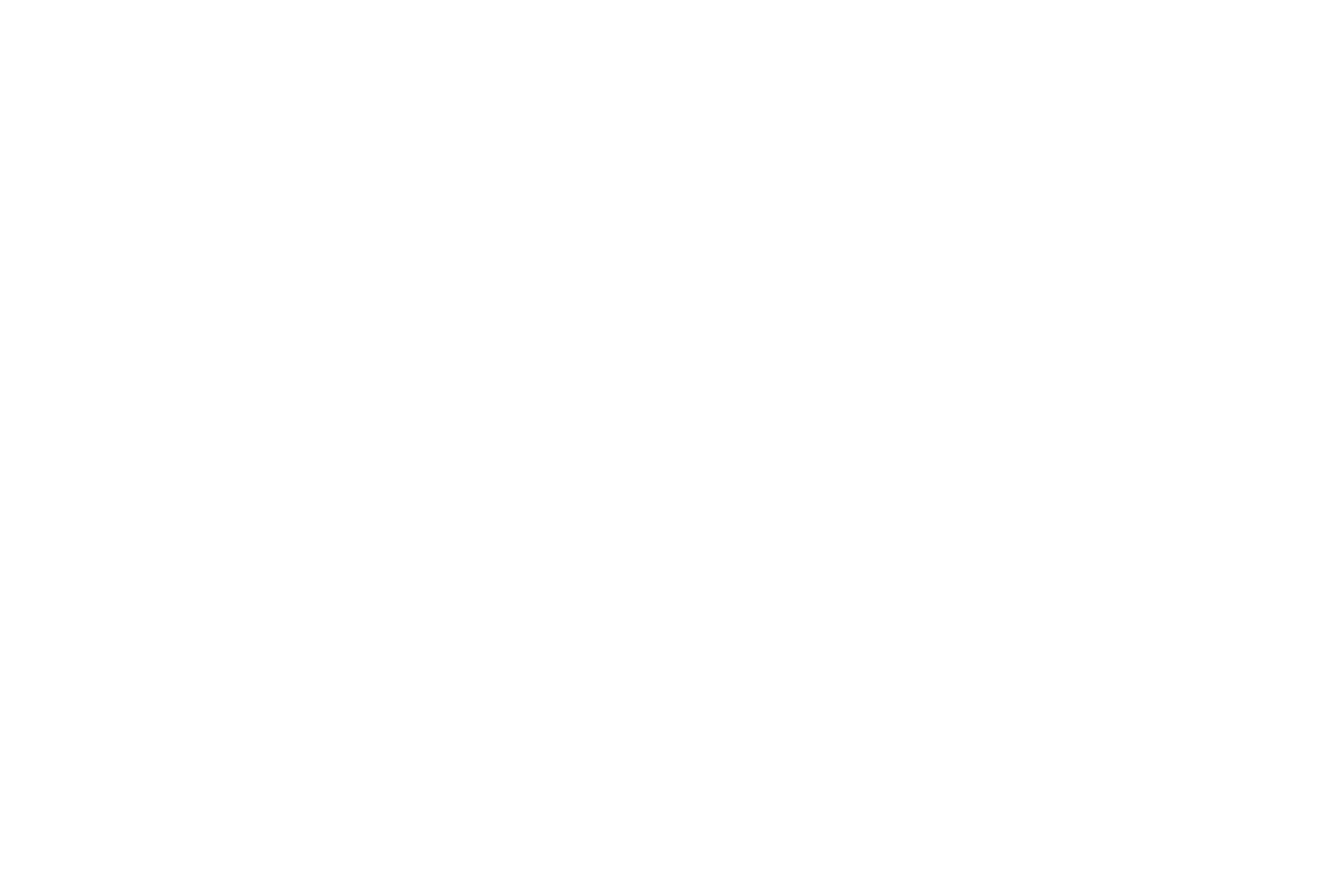}
\end{subfigure}
\caption{\label{fig:secho_data} \tonecth values measured for each sample. }
\end{figure*}

Solid echo data for hyperpolarized and thermal signals of each sample is given in Figure \ref{fig:secho_data}. The 95\% confidence interval of each biexponential model fit from the real component of the hyperpolarized data is underlaid in the figure. The thermal data are fit to this model with a scaling factor to estimate the amplitude of the thermal signal. Bootstrapping is used in order to gain proper estimates of the mean and variance of the amplitude of the thermal signal fitted from the hyperpolarized model data. Bootstrapping is carried out by saving thermal data sets in sets of 4 averages. Some thermal measurements require 1000-2000 datasets of 4 averages. These are randomly resampled with replacement approximately 10$^4$ times, where upon each resampling the data are averaged and a mean amplitude is fit. This produces a distribution of 10$^4$ fitted amplitudes whose mean and variance give a proper measure of the amplitude of the thermal \cthirt NMR signal. The number of times to resample was chosen such that the values given for mean and variance are reproduced, despite using random resampling. Because acquiring thermal signal in some cases involves thousands of averages, the solid echo pulse sequences for these experiments are run with a recovery time $t_{rec}$ of approximately \tonecth for efficiency. The amplitudes are then corrected with measured \tonecth data to obtain the true amplitude of the thermal signal. These correction factors are given in the subcaptions of Figure \ref{fig:secho_data}.



It is clear from the D1 solid echo data in Figure \ref{fig:secho_data}(a) the thermal \cthirt NMR signal is still dominated by noise. This measurement at least allows an assignment of a mean enhancement with upper and lower bounds to 95\% confidence. While the errors in the amplitude of the thermal relative to the hyperpolarized \cthirt NMR signal for D1 are normally distributed, the distribution of noise in the enhancement is not, so we cannot state a confidence interval in terms of 2$\sigma$. We instead find the bounds on the estimated relative amplitude of the thermal signal and convert these to bounds in enhancement. In the case of the \cthirt-enriched samples, \cthirt NMR signal is detected and therefore the errors in enhancement are normally distributed. For these samples, we fit the mean enhancement by bootstrapping the thermal data and fitting to the hyperpolarized model data, and the variance of this enhancement gives us the confidence interval. Estimated values for relative amplitude of thermal \cthirt NMR signals, final polarization in the hyperpolarized signal, as well as enhancements are given in Table S\ref{tab:char}.

\begin{figure*}[h!]
\centering
\includegraphics[width=0.35\textwidth]{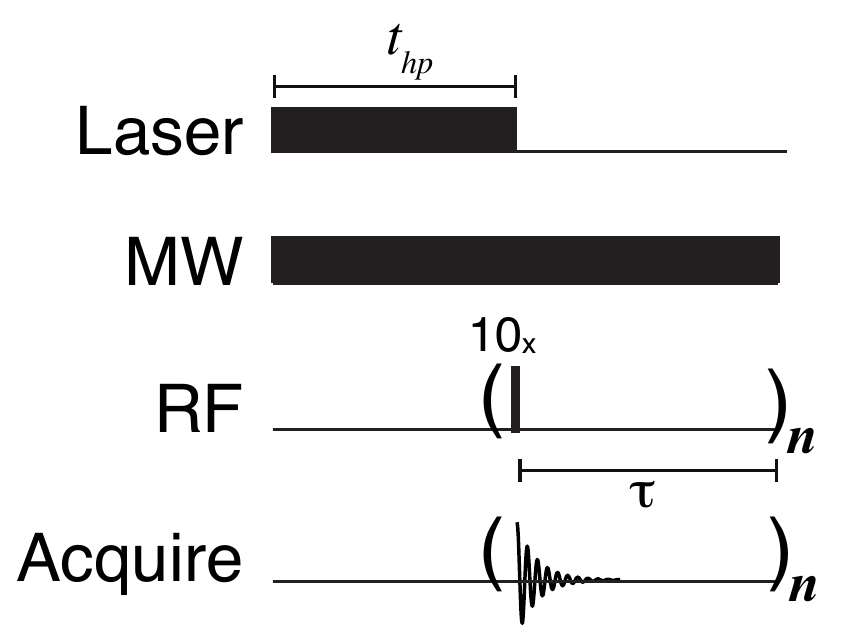}
\caption{\label{fig:sflipt1} Schematic of the small-flip angle \tone measurement. The \cthirt nuclei are hyperpolarized for a period of time $t_{hp}$ by applying the laser and microwave excitation at the optimal saturation frequency. The laser is then switched off with an optical shutter and a series of $n$ small-flip pulses along x are acquired spaced by a period $\tau$. The free-induction decay is acquired after each small-flip pulse. For clarity, 10\degree is chosen in the schematic. The microwaves remain on during the experiment in an effort to minimize differences in NMR signal due to heating. }
\end{figure*}

With the exception of the sample with 25\% \cthirt enrichment, the nuclear \tone measurements are carried out with a small-flip angle pulse sequence, described schematically in Figure \ref{fig:sflipt1}. The angle of the small-flip pulse is calibrated from a nutation curve and the final \tone value measured is corrected for the contribution of the small-flip pulse to relaxation \cite{li2008}. We use small-flip angle pulse lengths of 0.5 - $\SI{1.5}{\mu s}$ with a typical spacing $\tau$ of 1 s. For each \tone measurement the polarization is allowed to build for approximately 2\tdnp. All \tonecth measurements are given in Figure \ref{fig:allt1}.

\end{document}